\documentclass[12pt]{article}
\usepackage{epsfig}
\usepackage{psfrag}
\usepackage{amsmath}
\usepackage{lscape}
\usepackage{multirow}
\usepackage[super, comma, sort&compress]{natbib}
\newcommand{\bfig}{\begin{figure}[h] \begin{center}}
\newcommand{\efig}{\end{center} \end{figure}}

\begin{document}

\begin{center}
Steady--state simulations using weighted ensemble path sampling
\end{center}

\begin{center}
Divesh Bhatt, Bin W. Zhang, 
and Daniel M. Zuckerman\footnote{email: ddmmzz@pitt.edu}
\end{center}
\begin{center}
Department of Computational Biology,
University of Pittsburgh
\end{center}

\centerline{\Large \bf Abstract}

 We extend the weighted ensemble (WE) path sampling method to
perform rigorous statistical sampling for systems at steady state. 
The straightforward steady--state implementation of WE is directly
practical for simple landscapes, but not when significant metastable
intermediates states are present. We therefore develop an enhanced WE
scheme, building on existing ideas, which accelerates attainment of
steady state in complex systems. We apply both WE approaches
to several model systems confirming their correctness and efficiency by
comparison with brute--force results. The enhanced version is significantly 
faster than the brute force and straightforward WE for systems
with WE bins that accurately reflect the reaction coordinate(s).
The new WE methods can also be applied to equilibrium sampling, since
equilibrium is a steady state.

\section{Introduction}

 Steady--state phenomena are ubiquitous in  biological and
chemical systems. In biological systems,
this is rather unsurprising due to the approximate homeostasis observed 
at multiple scales (population level, cellular level,
and molecular level).  At a molecular level, homeostasis results in
a protein being constantly removed (for ``consumption'' downstream) 
after formation, and new protein is formed at a constant rate. 
Enzymatic reactions\cite{mcb,fersht}
inside the cell occur at a steady state if the concentration of reactants
remains constant and the product is continually removed (as is the
case under normal homeostatic conditions). The movement of motor
proteins\cite{mcb,pb}
under saturating ATP conditions typical of the cell is also well described
by steady state. 
Additionally, some {\it in vitro} experiments are
run at steady state -- for example, studies of enzymatic reactions
described by Michaelis--Menten kinetics.\cite{pb,fersht}

From a theoretical perspective, the steady stare is not simply a
``special case'' but is fundamentally connected to other ensembles.
First and most obviously, equilibrium is itself a steady state. But more
interestingly, the trajectories exhibited in a steady state are
identical to those which would occur in a single--molecule
or first--passage scenario. This was formalized by Hill,\cite{hill2}
who showed that the flux of probability from state A to B in
a steady state is exactly equal to the inverse mean--first--passage time
(MFPT) from A to B if the trajectories arriving at B are immediately
fed back into A. (As we recently showed, equilibrium can also
be exactly decomposed into two steady states\cite{symm}).
In this manuscript, we present a path sampling procedure
that establishes steady state efficiently and allows for the calculation
of both steady state and first passage rates.
 
 Computationally, various paths sampling procedures, such as Transition
Path Sampling,\cite{tps1,tps2,tps3,tps4} 
Transition Interface Sampling,\cite{tis1,tis2,tis3}
Forward Flux Sampling,\cite{ffs1,ffs2,ffs3,ffs4,ffs5}
and Milestoning\cite{mile1,mile2}
have been developed to compute rate constants for systems with slow
kinetics. However, to our knowledge these methods have not been applied to 
steady state. Recently, Dinner and coworkers\cite{dinner1,dinner2}
showed that the basic definition
of steady state (net zero flux in any region of configurational space) can
be used to establish steady state computationally. This procedure is an
analog of umbrella sampling for systems not at equilibrium: trajectories
in small divisions of configurational space are simulated independently
and net zero flux is enforced by monitoring the crossings of trajectories
to and from the neighboring regions.
Later, vanden--Eijnden and Venturoli\cite{van1} 
extended this procedure to allow the
calculation of both the steady--state rate and the steady--state flux.

 In this work, we extend the previously developed weighted ensemble (WE) path
sampling procedure\cite{huber,we_ass,bin} 
to perform rigorous statistical sampling of
systems at steady state. WE is a particularly attractive path sampling
procedure due to its simplicity. In contrast to Dinner and 
coworkers,\cite{dinner1,dinner2}
we do not enforce a strict net zero flux in a region. Rather, statistical
net zero flux emerges naturally at steady state from WE combined with a simple
feedback loop. In essence, net zero flux serves to determine whether
the system has reached a steady state. 

 More importantly, we also develop a probability adjustment procedure 
to enhance the attainment of steady state using WE simulation. 
Such a procedure becomes 
particularly important in systems that show significant intermediates
between the two end states, as expected for large biomolecules. 
Without such a probability adjustment, the
steady state is achieved only slowly due to the waiting times required for the
regions with the intermediates to reach a stationary value of the probability.

 This manuscript is organized as follows. First we show how steady state
can be obtained in rigorous statistical simulations using a feedback loop
and introduce our method as applied to WE path sampling. Then, we describe
in detail the enhanced probability adjustment procedure that leads to a more
efficient establishment of steady state. We also describe the different
model systems we use: one-- and two--dimensional toy systems, and all--atom
alanine dipeptide. Following that, we present results for these systems where
we also compare the results with brute--force simulations where possible.
This is followed by a discussion of the computational issues and possible
applications and improvements. Finally, we present our conclusions.

\section{Method}
\label{method}

\subsection{General points}

 For a system in steady state, the probability of visiting a part
of the configuration space remains constant:
\begin{equation}
\frac{dP_i}{dt}=\sum_j f_{ji} -\sum_j f_{ij} = 0
\label{e0}
\end{equation}
where $P_i$ is the probability of an arbitrary region $i$, and $f_{ji}$ is the
flux of probability from region $j$ to region $i$. 
Together, the regions $i$ and the set $\{ j\}$ completely cover the space
but do not overlap. Eq~\ref{e0} says that
the net flux into
a region equals the net flux out of the region.
As such, equilibrium is a special steady state, where $f_{ij}=f_{ji}=0$
for all pairs of regions.
A more general steady state can be obtained even with flows -- {\it i.e.},
with $f_{ij}\ne 0$.

 A common steady state is obtained by a feedback loop from the final
state to the starting state -- if a system reaches the final state,
it is fed back into the initial state. This type of steady
state is prevalent in biological systems: the proteins formed in the
nucleus are removed (for use in cellular processes), subsequently break down,
and the amino acid residues are fed back for protein synthesis by the DNA.
A more direct feedback loop is observed in case of enzyme catalysis -- the
enzyme is ``fed back'' to catalyze more reactants to products (assuming a
constant reactant concentration -- either under homeostatic conditions in
organisms, or via an external reservoir in chemical systems).

 These two simple ideas -- that steady state is established via a feedback
loop, and is obtained when net probability fluxes
are zero -- are used to develop methods to attain steady states,
building on earlier work.\cite{dinner1}

\subsection{Steady state with brute force}

 Before proceeding to a detailed discussion of simulating a steady state via WE,
we note that a steady state can be established using brute--force
simulations via a simple feedback
loop from the end state to the initial state. If a large number of
(independent) brute--force trajectories are started 
from the initial state, then
such a feedback loop eventually establishes a steady state -- the
distribution of trajectories at any part of the configurational space
is independent of time.

For systems with large barriers between the two states,
brute--force simulations are not very efficient to study
transitions from one state to the other.

\subsection{Brief review of generic Weighted Ensemble}

 Weighted ensemble path sampling is described in greater detail in the
original paper by Huber and Kim,\cite{huber} 
and a more recent theoretical review.\cite{bin2}
Here we give a brief overview of ``ordinary'' WE path sampling
before discussing our WE methods for steady state.
Typically, ``start'' and ``end'' states are defined in advance.
Further, the whole configuration space is divided into
bins, and several trajectories typically are started from one bin with each
assigned an equal probability. These trajectories
are allowed to evolve for a certain time increment, $\tau$, using
the natural system dynamics. After each $\tau$, the trajectories are
checked to determine the occupied bins. Each time
a bin is occupied, the trajectories entering the bin are 
split or combined to give a predetermined number of trajectories for that
bin. Trajectory probabilities are accordingly allotted in a 
rigorous statistical
manner. That is, no bias is introduced in the evolution of the system, and
each part of the configuration space retains the correct probability (as
required by the natural, unbiased system dynamics) at all times.
As in a Fokker--Planck picture, the system evolution
depends solely on the initial condition.\cite{bin2}

\subsection{`Regular' weighted ensemble attainment of steady state}

 Weighted ensemble path sampling is naturally adapted to simulating 
a steady state: any trajectory that enters the end state is fed back 
into the starting state. Naturally, this requires definitions of the starting
and the target states. The starting state can be a finite region of the
configuration space or a single configuration (for example, a fully extended
protein conformation). The target state must be a finite region of the
configuration space to enable a finite flux into the target state. In
some cases, the target state may be known in advance -- and we consider
such systems in this work.

Since the dynamics are not perturbed at all in WE simulations, 
each point in the
configuration space occurs with the correct probability for the
condition simulated (whether at
equilibrium, in steady state, or for an evolving distribution).\cite{bin2}
Correspondingly, because a feedback loop is part of the steady--state
definition, it does not perturb the steady state system.

\subsection{`Enhanced' weighted ensemble attainment of steady state}
\label{ewem}

In the presence of one or more significant (metastable) intermediates,
a steady state can only be achieved after the
intermediate state is substantially populated. In such cases, regular
WE simulation is still quite inefficient due to the ``relaxation''
time required from the initial conditions (whereas the steady state solution is
completely determined by the feedback condition). This relaxation time
can be eliminated if the initial conditions are chosen to approximate
the steady state probability distribution.

 As a simple example, consider a system with 4 discrete states.
State 1 is the starting state and state 4 is the target state. In case
of Markovian dynamics, the rate of change of probability in any state
is exactly described by the reformulation of eq~\ref{e0}:
\begin{equation}
\frac{dP_i}{dt}=\sum_j k_{ji}P_j-\sum_jk_{ij}P_i.
\label{e1}
\end{equation}
where $k_{ij}$ is the rate of transition from state $i$ to state $j$.
For simplicity of the following discussion, we assume that all 
$k_{ij}$ are equal. At steady state, $P_4\equiv 0$, and the solution
of eq~\ref{e1} gives $\{ P_1,P_2,P_3,P_4\} = \{ 1/2,1/3,1/6,0\}$.
A procedure that utilizes an initial condition close to $\{1/2,1/3,1/6,0\}$
effectively eliminates any relaxation time to steady state, whereas the initial
conditions of $\{1,0,0,0\}$ may require a substantial relaxation time,
depending on the magnitude of the rates.

 Motivated by the above discussion and previous work,\cite{dinner1} 
we utilize eq~\ref{e1} to estimate
the steady state probabilities in regions (or bins) of the 
configurational space. The transition matrix elements, $k_{ij}$, are
determined by running unbiased dynamics on trajectories in a bin $i$
for a time increment $\tau$ and estimating the conditional probability
to transition to bin $j$.
Because generic WE always uses unbiased dynamics, these transition matrix
elements can, for example, be computed by setting up a short WE run
from arbitrary initial conditions -- we elaborate upon this shortly.
From the obtained values of $k_{ij}$, estimates for
$P_i$ are obtained via the
use of eq~\ref{e1}. Specifically, if B denotes the target, and A
denotes the initial bin, then eq~\ref{e1} is used for all bins
except A and B. For the target bin, $P_{\mathrm{B}}=0$, and the sum
of all bin probabilities is 1 ({\it i.e.}, 
$P_{\mathrm{A}}+\sum_{i\ne\mathrm{A}}P_i=1$). 
With $N$ bins, this system of $N-1$ linear
algebraic equations with $N-1$ unknowns ($P_{\mathrm{B}}=0$) 
can be solved by standard methods such as Gauss Elimination.

 The obtained values of probabilities in each bin 
thus obtained constitute a near steady--state initial condition.
The system is then allowed
to relax using regular weighted ensemble until steady values
are obtained. To the extent that the estimates of bin probabilities 
are accurate, the relaxation time is minimized.

 There are two important technical issues that determine the appropriateness
of the calculated transition matrix of rates for use in eq~\ref{e1} to
determine $P_i$ values. One is
related to the Markovian approximation inherent in eq~\ref{e1}:
ideally, the rate $k_{ij}$ for transitions from bin $i$ to bin $j$ 
should be calculated only
after the intra--bin relaxation time, $\tau_i$ ({\it i.e.},
$\tau >\tau_i$). This issue of the applicability of the rate equation
was also discussed by Buchete and Hummer.\cite{hummer2}
Furthermore, several transitions from bin $i$ must 
be observed to obtain
good statistics. Clearly, a longer wait time between recording
transitions, as well as a studying larger number of transitions
will give a more accurate estimate of the transition matrix elements
at the cost of increasingly longer time to calculate the
steady state distribution. However, we find that a reasonable estimate of
the steady state distribution via the transition matrix
in a short amount of time (compared to establishment of steady state
via regular WE) can still be obtained.
Some of the complications just discussed are artifacts of our
(simple) implementation: we use a single WE $\tau$ interval for $k_{ij}$
estimate, but better ``book keeping'' will readily permit the use of
multiple $\tau$.

 The arbitrary initial conditions for computing rate matrix elements
can be delta function at the starting state, or obtained via running
dynamics at elevated temperatures to ``fill'' up the space initially.
Further, after the initial conditions for rate matrix computation are
obtained, short brute--force dynamics from these initial conditions
may repeatedly be performed to obtain statistically meaningful rates.
The only requirement for an appropriate estimate of the rate matrix elements
is that dynamics be unbiased at the temperature of interest while computing
the rates.

\subsection{Equilibrium sampling}

 The procedure outlined for enhanced WE attainment can also be utilized for
performing equilibrium sampling. Instead of setting the
probability of the target bin to zero, we use the expression for
$dP_{\mathrm{B}}/dt=0$ from eq~\ref{e1} (as for every other bin). 
This system of $N$ linear algebraic equations with $N$ unknowns can, again, be
solved using standard methods. An accurate estimate of bin populations
gives the exact equilibrium distribution. In most
problems, this requires a ``relaxation'' to equilibrium.
However, a good estimate can dramatically reduce the
relaxation time. We mention that the use of rates to 
estimate the equilibrium probability distribution was also explored by 
Buchete and Hummer\cite{hummer2} and by Pande and coworkers.\cite{pande1}

\subsection{Analysis of WE efficiency versus brute force to obtain 
steady state}
\label{eff}

 There is an intimate connection between steady--state and first--passage
kinetics, which permits a fairly direct comparison between our WE
simulations and traditional brute--force simulations. In particular,
as described in the Introduction, Hill\cite{hill2} showed that the probability
arriving per unit time ({\it i.e.}, the flux) into the `end' state
in a steady state with immediate feedback in exactly equal to the
`traditional' rate or inverse mean-first-passage time. In the
notation of eq~\ref{e0}, this can be written as
\begin{equation}
\sum_{i\ne \mathrm{B}} f_{i\mathrm{B}}(\mathrm{SS})=k_{\mathrm{AB}}=
 \frac{1}{\mathrm{MFPT\;A}\rightarrow\mathrm{B}},
\label{e3}
\end{equation}
where SS denotes that the fluxes $f_{i\mathrm{B}}$ occur in steady state.
Our WE simulations yield all the steady state fluxes $f_{ij}$(SS),
which enable quite precise estimates of the brute--force rate
$k_{\mathrm{AB}}$.

 There are several different measures for comparing the efficiency
of enhanced WE path sampling to brute--force simulations. One measure
is the comparative transient time that these methods require to first reach
within a certain precision of the steady--state flux. Another measure
is the comparative time that enhanced WE and brute--force simulations
require to attain a certain level of precision {\it after} the transients
have decayed. A final comparison is via the time either method takes
to ``find'' a pathway between the initial and the final state.

 We mainly focus on the first of these measures to compare the
efficiency of enhanced WE to brute--force simulations: the
comparative time required by these two methods to first reach
within a certain precision of the steady--state flux.
The reason for this choice is that although we will explicitly obtain 
both timescales (first approach to steady--state flux, as well as 
sampling after transients) for enhanced WE simulations, we
only estimate one timescale -- MFPT from eq~\ref{e3} -- 
of relevance for the brute--force
simulations. As we discuss below further, MFPT most naturally gives
the approach of brute--force simulations to steady--state flux.

 To quantify this comparison, we first 
determine the steady--state flux and the standard deviation via block 
averaging\cite{block} obtained from enhanced WE simulations after the
transients. Next, we determine the total time required for
the block averaged flux to first reach within a one standard deviation of
the mean steady state flux.
This total time includes the initial phase during rate
computation. We thus obtain the total time, $\tau_{\mathrm{tot}}$ 
that each trajectory in
enhanced WE simulations has been ``active'' for before steady state is
attained. The number of trajectories multiplied by
$\tau_{\mathrm{tot}}$ gives the total simulation time for enhanced
WE simulation to first reach within one standard deviation of the
mean steady--state flux.

 The time required for brute--force simulations to first attain the same
level of precision can be estimated from MFPT as follows.
Assuming a Poissonian process for hopping between two 
states separated by a high barrier, the distribution of first passage 
time from the initial to the final state depends upon the number of
intermediate basins leading to a distribution that is approximated
by a gamma distribution\cite{gamma1} determined by the number of 
intermediate states. (A gamma distribution is a convolution of simple
exponential distributions).
For example, if there are $n$ intermediate basins along a pathway
from the initial to the final state, and if the rates of transition among
all consecutive bins are equal ($\equiv\kappa$), the MFPT is given by
$n\kappa^{-1}$, and the variance in MFPT by $n\kappa^{-2}$. 
For $N$ brute--force
trajectories, this leads to expressing one standard deviation from
the mean by $\sqrt{k\theta^2/N}$. Equating the MFPT and the variance
obtained from enhanced WE to the model gives an estimate of $N$: if $N$ 
brute--force simulations are performed for a total of MFPT, the flux into 
the final state will be within one standard deviation of the mean. 
And, $N\times$MFPT gives an estimate of the brute--force simulation time
required to get within one standard error of the mean steady--state flux.

 Returning to other methods for comparing the efficiencies of
enhanced WE and brute--force simulations,
 we note that WE is significantly faster than brute--force
simulations in determining transition paths for a variety of
systems.\cite{bin,bin2}

\section{Model systems and dynamics}

\subsection{One dimensional toy model}

We start with the simplest model that allows us to establish
some of the concepts discussed above regarding high barriers
between states and significant intermediates. For this purpose,
we study the one--dimensional potential energy function 
shown in Figure~\ref{pot1d}. The system has
two three energy wells, and the barriers  between the well minima are
$6 k_{\mathrm{B}}T$. As usual, $k_{\mathrm{B}}$ is Boltzmann constant and
$T$ is temperature. The presence of the intermediate state and the 
high energy barrier
makes the system challenging for brute--force and regular WE
simulations. Full details of potential are given in the Appendix.

The target state (state B) is defined as
$x>4.5$. The initial state (state A) is delimited by the two
dashed lines on the left in Figure~\ref{pot1d} 
($0.98 <x<1.17$). The probabilities of
trajectories that are fed back into state A after entering the 
target state are distributed proportionate to
the probabilities of the existing trajectories in state A. 
In this example for WE simulations, the
abscissa is divided into 25 bins with the first bin for $x<0$,
and the last bin for $x>4.5$ (the target state B). The other 23
bins are equal width in between these two bins.

 We used the overdamped Langevin equation with the following
discretized version:
\begin{equation}
x_{j+1}=x_j-\frac{\Delta t}{m\gamma}\left( \frac{dU}{dx}\right)_{x_j}
 + \Delta x^{\mathrm{rand}}
\label{e2}
\end{equation}
where $x_j$ is the one--dimensional coordinate,
$\Delta t$ is the time
step for integration, $m$ is the mass of the particle,
$\gamma$ is the friction constant with units of s$^{-1}$. The term
$\Delta x^{\mathrm{rand}}$ is the random displacement (modeling collisions)
chosen from a
Gaussian with zero mean and a variance of $2kT\Delta t/(m\gamma)$.
We selected $2kT\Delta t/m\gamma = 0.001$ m$^2$ 
and each increment, $\tau$, for WE simulation corresponds to 10 
such integration steps.

\subsection{Two dimensional toy model}

 One dimensional models are intrinsically limited as test systems, however: 
the binning and landscape dimensionalties
are the same, and only one pathway between the initial
and the target state is possible. 
For more realistic systems, this
does not hold true. A two--dimensional system is the simplest
system that allows us to relax these limitations and also
allows for testing the methods for less than optimal reaction
coordinates. Accordingly, we consider two--dimensional potential energy 
surfaces shown in Figures~\ref{2d_test1} and \ref{2d_test}, each of which
possesses two reaction channels separated by significant barriers.
The potential energy surface of Figure~\ref{2d_test1} is
less rugged than that of Figure~\ref{2d_test}. (In either
figure, the two panels represent different binnings --
discussed shortly -- for the same potential energy surface).
Individual wells and the barriers are modeled by two--dimensional Gaussians 
and are deeper for the more rugged potential energy surface of 
Figure~\ref{2d_test}.

 In both the figures, 
the two end states are labeled, and any trajectory that enters
state B is fed back to a single point at $x=-3.5$ and $y=-0.5$

We employ two different types of binnings for each surface
for WE simulations, as shown in Figures~\ref{2d_test1} and \ref{2d_test}.
A two dimensional binning allows for only one potential energy well
in a bin so that bins match the reaction coordinate. 
On the other hand, one dimensional binning is not optimal because
well--separated potential
energy wells are present in one bin.
Depending upon the heights of the barriers between the wells,
1D binning could requires significant transverse relaxation time
within a bin. 
As a consequence, the estimate of bin populations obtained 
via eq~\ref{e1} may be less accurate -- requiring a longer ``relaxation'' 
time to steady state.

 For notational ease, we refer to a particular bin by its ``index''.
For example, the center of State A in both Figures~\ref{2d_test1} (a)
and \ref{2d_test} (a) has an index (2,7) -- implying that the bin which
is in the second discretized region along the $x$ coordinate and
the seventh discretized region along the $y$ coordinate contains the
center of State A.

 We used a two--dimensional overdamped Langevin equation  with the
$y$ coordinate governed by the analog of eq~\ref{e2}. The
parameters of the Langevin equation are exactly the same as that for
the one--dimensional system.

\subsection{Alanine Dipeptide}

We study all-atom ``alanine dipeptide'' (or
alanine with acetaldehyde and n-methylamide capping groups).
Although alanine dipeptide is a relatively small biomolecule,
it has a complex energy landscape characterized by intermediates and
multiple pathways. At the same time, the paths can be visualized
readily in a small set of coordinates and thus alanine dipeptide has
been a frequent target for path sampling.\cite{adp1,adp2,adp3,adp4}
The significant intermediates present in implicitly or explicitly
solvated alanine dipeptide make it a good challenge for steady--state
studies. Our results will bear out the challenges in this small molecule.

 For alanine dipeptide in implicit solvent, the configurational
space can effectively be condensed to a two--dimensional
space given by the $\psi$ and $\phi$ torsional angles.
In Figure~\ref{adp_pot}, 
exploratory brute--force simulations show regions of this
two--dimensional configurational space that are populated
significantly with a high density of red dots. Also shown in
the figure are dashed circles approximately representing states. The region
labeled C$_{7\mathrm{eq}}$ contains the starting state
(a delta function: $\Phi =-77.9$ and $\Psi =138.4$), 
and C$_{7\mathrm{ax}}$ is the target state (within a radius of 20 degrees
about $\Phi=61.4$ and $\Psi=-71.4$).
There are two significant intermediate states -- labeled
$\mathrm{\alpha_R}$ (right--handed helix) and
$\mathrm{\alpha_L}$ (left--handed helix). The lines show the
WE binning in the $\Phi$ and $\Psi$ directions.
Note that circularity/periodicity of the $\Phi$ and $\Psi$
angles engenders a multiplicity of pathways between any two states.

An fully atomistic alanine dipeptide molecule
using CHARMM 19 forcefield  with implicit solvent ACE (analytic
continuum electrostatics) model\cite{charmm1} is simulated via Langevin
dynamics with integration step of 1 fs. The friction constant is set to 
50~s$^{-1}$.
The solvent used is the Analytic Continuum Electrostatics model with
the dielectric constant in the region occupied by explicitly modeled
atoms as 1 (IEPS parameter), the dielectric constant of space occupied
by the solvent as 80 (SEPS parameter), the Gaussian width of density 
distributions describing volumes of atoms as 1.3 (ALPHA parameter), and
the hydrophobic scaling factor, SIGMA, as 3.0. Further, atom--based
switching is used.

 WE simulations were run as described in Section~\ref{method},
with the $30^{\circ}\times 30^{\circ}$ bins shown in Figure~\ref{adp_pot}.
 After the probability adjustment in enhanced WE, 
the weighted ensemble trajectories
are split and combined every 100 fs to keep 20 trajectories
in each occupied bin. At these intervals, the trajectories
are also analyzed to check if any reach the target state.
However, 100 fs interval was determined to be too short for
an accurate estimate of the bin populations using eq~\ref{e1}.
Thus, for the transition matrix evaluation, the trajectories
are checked for transitions every 1000 fs for 500 K, 
and every 10000 fs for 300 K.
Our WE simulation run CHARMM via an in--house C program available
to the public.

\section{Results}
\label{result}

\subsection{One--dimensional system}

 First, we show the correctness of the WE methods and the efficiency
gain in the enhanced WE on the one--dimensional energy landscape of
Figure~\ref{pot1d}. The system is challenging because of its
high barriers and a deep intermediate state.
At steady state, all bins must have time--independent probabilities,
and the target state must have a constant probability flux.
Figure~\ref{flux1d} shows the flux into state B 
as a function of time obtained
using both regular and enhanced steady state procedure of the
weighted ensemble path sampling. The two methods agree well - reflecting
the fact that the adjustment of initial conditions in the enhanced WE does not
affect the steady state solution. Moreover, it shows that
the probability adjustment procedure does not disturb the
natural system dynamics.

 Most importantly, Figure~\ref{flux1d} shows that 
enhanced WE simulation does indeed achieve the steady--state 
flux value significantly
quicker than regular WE. For such a simple system, as used
in this example, it is not surprising that the probability adjustment
procedure based described above gives a reasonably accurate
estimate of the bin probabilities.
However, more fundamentally, systems with significant intermediates
are not optimally studied via standard WE -- nor by, presumably,
other methods which fail to ``shortcut'' intermediate dwell times.

\subsection{Two--dimensional system}
\label{2d}

 The two--dimensional systems of Figure~\ref{2d_test1} and \ref{2d_test}
permit us to examine the performance of WE in meeting additional
challenges: multiple intermediates, more than one reaction channel, and
suboptimal bins.

 Again, we first check the correctness of
our WE methods for steady state simulation. 
We plot the fluxes into state B for the smooth and rough two--dimensional
potential energy surfaces in Figures~\ref{flux2d_4} and \ref{flux2d},
respectively. Each figure show results obtained using both one-- and 
two--dimensional binnings (see Figures~\ref{2d_test1} and \ref{2d_test}). 
For the less rugged energy landscape, it is possible to obtain accurate
brute force results for the mean first passage time, and the inverse
of MFPT is the steady state flux into state B. Clearly, this agrees
with the steady state flux values obtained using both the enhanced
and the regular versions of WE (and with either binning procedure).

 One very important observation from Figures~\ref{flux2d_4} and \ref{flux2d}
is that the it takes approximately the same amount of time to reach the
steady state using enhanced WE simulation, irrespective of the ruggedness of
the landscape. However, this is clearly not true for either regular WE
or brute--force simulations where the rate decreases by more than three
orders of magnitude for the rugged landscape. This again
emphasizes the utility of the probability adjustment procedure.

One--dimensional binnings (for both the enhanced and regular
WE simulations -- red line and the magenta line, respectively) also lead 
to the correct steady--state flux values into state B. However, the
data are noisier. Despite the noisier data, it is clear that the
enhanced version reaches the correct steady state flux value faster
even for the one--dimensional binning.

 Further insight into the discrepancy between regular and
enhanced WE simulations for the more rugged landscape is obtained by
examining an individual intermediate state. Figure~\ref{pb2d} shows
the population of a single bin at index (10,4)  using
both the enhanced (black line) and the regular (blue line) version.
Clearly, the enhanced version reaches a plateau value fairly quickly.
On the other hand, the population of this bin using regular
WE does not reach a plateau value for 10$^5$ $\tau$ increments, and
is still two orders of magnitude lower than the plateau value
of the enhanced version. Thus, Figure~\ref{pb2d} clearly illustrates
that regular WE attains steady state via regular WE very slowly if
there are significant intermediate states.

\subsubsection*{Efficiency comparison with brute force}

 To further quantify the gain due to the enhancement procedure,
 we explicitly compare the efficiency of the enhanced WE
approach with the brute--force simulations using the procedure
discussed in Section~\ref{eff}.
For both the potential energy surfaces of Figures~\ref{2d_test1}
and \ref{2d_test}, there are three or four intermediate states, 
depending upon the path. We use $n=3$ for the number of intermediate
basins (using another similar number does not qualitatively affect
the comparison below). Further, we compare only for the two--dimensional
binnings. 

 For the smoother potential energy surface of Figure~\ref{2d_test1}, the
total simulation time required by enhanced WE to reach within one standard
error of the steady--state flux is 2$\times 10^7\tau$. On the other hand,
the estimate for brute--force simulations is 4$\times 10^5\tau$. Clearly,
for this relatively smooth landscape, brute--force simulations are more
efficient than enhanced (and regular) WE.

 On the other hand, the converse is obtained for the rougher landscape
of Figure~\ref{2d_test}. Whereas the enhanced WE requires a similar aggregate
time to reach the desired deviation from the steady--state flux, the
estimate for brute--force simulation balloons to 4$\times 10^8\tau$. Thus, with
an increase in the roughness of the energy landscape, the enhanced WE 
progressively outperforms brute--force simulations. This is expected
since the time required for brute--force simulations increase in proportion
to the MFPT, whereas the enhanced WE requires only a few transition events
for the probability adjustment to steady state because WE always
finds the transition paths rapidly.

\subsection{Alanine dipeptide}

Again, for alanine dipeptide, we first demonstrate the correctness of the WE
methods by comparing with independent brute--force estimates, and 
then describe
the efficiency gain from the use of enhanced WE. In addition to $T=300$~K,
an elevated value (500~K) is studied to permit quantitative comparison
with brute--force simulations.

Figure~\ref{adp_500} plots the flux into the target
state at 500 K using both enhanced and regular versions of WE, along
with an independent estimate of flux from brute--force simulations.
Both versions give the correct steady state flux values, and these
values are obtained in approximately the same number of
$\tau$ increments at $T=500$~K.

 A similar plot for $T=300$~K is given in Figure~\ref{adp_300} (a).
As expected, the flux into
state B is lower at 300 K compared to 500 K. More importantly,
the enhanced version of WE agrees with the brute--force results,
however, the flux into state B obtained using the regular version
has not reached the correct steady state value in 1400 time increments.
Figure~\ref{adp_300} (b) further emphasizes the difference between
the enhanced and the regular version: the probability in a particular
bin with index (4,5) which contains an intermediate state 
has clearly reached a plateau value in enhanced WE
and not in regular WE.

\subsubsection*{Efficiency comparison with brute force}

 We compare the efficiency for enhanced WE against brute--force simulations
via the method discussed in Section~\ref{eff}. Figures~\ref{adp_500} and 
\ref{adp_300} give the steady--state flux and the variance. The gamma
distribution model we use has one intermediate basin ({\it i.e.}, $n=1$).
We chose $n=1$ since Figure~\ref{adp_pot} suggests one intermediate basin
along the path from the initial to the final state. However, using other
small values of $n$ does not change the qualitative comparison.

 We compute the total simulation time using both enhanced WE and
brute--force simulations to first attain steady state.
 At 500~K, the initial phase for rate computation lasts for 10$\tau$
and each $\tau$ during this phase is 1000~fs. This leads to a total
simulation time (including all the trajectories) to reach within 
one standard deviation of the mean steady--state flux
using enhanced WE as 60~ns. For $k=1$, brute--force
simulations is estimated to require approximately 70~ns. At 300~K,
a similar analysis yields an enhanced WE time of 150~ns and
a brute--force time of 5~$\mathrm{\mu}$s.
Clearly, enhanced WE method is significantly more efficient as the
temperature is decreased.

\subsection{Equilibrium sampling}

 As mentioned above, the WE steady--state method can easily be adapted to
equilibrium sampling, because equilibrium is a steady state.
Figure~\ref{equil} and \ref{equil_pot2} illustrate WE applied to 
equilibrium: they shows 
the populations of two different bins as a function of WE time obtained via
enhanced and regular WE simulations for the 2D potential energy
surfaces of Figures~\ref{2d_test1} and \ref{2d_test}, respectively. 
For the smooth potential energy surface, enhanced WE simulation reaches
equilibrium in a fairly small number of $\tau$ increments, whereas
regular WE requires more time to equilibrate as shown in Figure~\ref{equil}.
More dramatically, the regular version does not reach equilibrium
in the simulation timescale for the rugged potential energy landscape
as shown in Figure~\ref{equil_pot2}. On the other hand, even after
accounting for an initial time before probability adjustment
(equal to $2\times 10^{5}\tau$ for the rugged landscape and only 2000$\tau$
for the smoother one), the enhanced WE remains extremely efficient
when compared to regular WE simulations.

\subsubsection*{Efficiency comparison to brute--force simulations}

 We now compare the efficiency of enhanced WE to brute--force simulations.
From a knowledge of the MFPT for this system, we can also roughly estimate
the time required for achieving equilibrium via brute force simulations.
We expect equilibrium to be established via brute--force simulations
when the brute--force trajectories have traversed the region between
the initial and the final state several times. As in Section~\ref{2d},
we use the timescale provided by the MFPT and the number of
``equivalent'' brute--force trajectories ($n=10$, as in Section~\ref{2d})
to obtain a rough estimate of the time required for equilibration
via brute--force simulations.

For the smoother potential energy surface, $n\times$MFPT equals
$5\times 10^5\tau$, whereas it equals $4\times 10^8\tau$ for the
rugged potential energy surface. Establishment of equilibrium via brute--force
would require several times these values. On the other hand,
enhanced WE simulations require of the order of $5\times 10^7\tau$
($=1000\tau$ for establishment of equilibrium $\times 4500$ trajectories)
for establishing equilibrium for the smooth potential energy surface,
and approximately $5\times 10^8\tau$ for establishing equilibrium
for the rugged energy landscape.
Thus, the enhanced WE simulation becomes more efficient in establishing
equilibrium as the potential energy surface becomes more rugged.

\section{Discussion}

\subsection{Comparison with Markov models}
\label{markov}

 The use of eq~\ref{e0} is clearly reminiscent of Markov models of
the configuration space,\cite{wales,chodera}
and, for an exact Markovian decomposition, the probabilities obtained
via eq~\ref{e0} should be exact. However, the enhancement procedure we
use is qualitatively different from these Markovian models in two
crucial ways that we discuss below.

 First, we use eq~\ref{e0} only to estimate the steady--state probabilities
in order to shorten the relaxation time to the exact steady
state. Thus, the final rates that we obtain are exact and completely
independent of a Markovian approximation. Second, the paths obtained
via the procedure we presented above are continuous -- {\it i.e.}, we
generate trajectories that include time spent within a state. In contrast,
Markov models focus only on the hopping between Markovian states.

 Despite our approach not relying on Markov models, the efficiency of
our probability adjustment protocol for enhanced WE improves significantly
by considerations relevant to the Markov models.
To elaborate, we recall that the continuous--time trajectories 
mandate that the rates,
$k_{ij}$, used in eq~\ref{e0} be computed only after a certain
relaxation time within a bin for the Markovian picture to 
emerge.\cite{hummer2} Accordingly, the estimates of steady--state
probabilities using eq~\ref{e0} improve if the rates between
bins are computed after the relevant intrabin relaxation time.

\subsection{Comparing enhanced and regular WE simulation}

 The main reason for the efficiency gain obtained via
enhanced WE as compared to regular WE is that the enhanced WE
inherently utilizes the timescale inherent in ``fast'' trajectories 
to perform the probability adjustment.
In any stochastic process obeying the Fokker--Planck picture (and
WE clearly does that\cite{bin2}),
there is a distribution of transition times from the starting to
the target state. Although the nature of this distribution depends
upon the exact landscape, the ``fast'' transitions occur earlier than
the mean transition time. The rates between all pairs of bins are
computed fairly quickly after the fast trajectories reach the final
state, and the probability adjustment procedure of enhanced WE is applied.
On the other hand, regular WE must be applied for a time of the order
of MFPT to approach steady state.

 Due to the wait time for the probability adjustment procedure being
of the order of the time it takes the fast trajectories to reach the
target state, the efficiency gain for to enhanced WE as the energy 
landscape becomes more rugged becomes more pronounced (as is clear from a
comparison of Figures~\ref{flux2d_4} and \ref{flux2d}). The time it takes
for the fast trajectories to reach the target state is affected less than
the MFPT as the ruggedness of the landscape increases.

\subsection{Possible improvements}

 In this section, we discuss several possible ways
to improve the efficiency of the enhanced WE further.

 One main avenue for improvement is the construction of better
bins between initial and target states. As discussed in
Section~\ref{markov}, the relaxation time after probability adjustment
is minimized if the adjustment procedure results in each bin
displaying its steady--state probability. Such optimal bins may
be constructed from initial paths or fast trajectories between the 
starting and the target state. Further, construction of smaller bins
reduce the likelihood of significant transverse relaxation within
a bin -- resulting in a better estimate of steady--state
bin probabilities using eq~\ref{e0}.

 Even with somewhat suboptimal bins, it is possible to reduce the
relaxation time to steady state after probability adjustment by
appropriate choice of the number of trajectories in each bin.
Currently, all the bins are constructed to have the
same number of trajectories -- this may not be most efficient for relaxation to
steady state after adjustment of probabilities. Bins that are assigned
a high population by the probability adjustment procedure but show
a subsequent slow decrease in the bin population 
(due to small rates to other bins) 
are likely to show a faster relaxation upon an
increase in the number of particles. Thus, if the number of particles
in each bin is adjusted based on the rate calculation and the
assigned probability, the relaxation to steady state may be
significantly increased in the enhanced WE.

 Another possible strategy for improvement is the use of multiple
probability adjustment. As discussed in Section~\ref{ewem}, only if $\tau$
is longer than intrabin relaxation time that an appropriate
estimate of the interbin rates are obtained: the distribution of
trajectories within a bin reaches the appropriate steady--state distribution
only after the relaxation time. This relaxation time
depends on the initial distribution of trajectories, and is minimized
as that distribution approaches steady--state values. Accordingly,
subsequent instances of rate computation start from trajectories in
a bin that are increasingly closer to the steady--state distribution.
In this manner, progressively smaller $\tau$ can be used for each
probability adjustment segment -- thus, improving probability estimates
for a given simulation time.

 Further improvement is possible if the WE script is fully integrated
with the code for underlying dynamics. Currently, the overdamped
Langevin dynamics code for the toy models is ``hardwired'' into the
WE code, however, the WE script calls the Langevin dynamics in
CHARMM for alanine dipeptide for each trajectory at the beginning 
of each $\tau$ increment. This results in a significant overhead
associated with reinitialization of the underlying CHARMM code in
each instance. Due to this, an increase in the magnitude of each
$\tau$ does not scale linearly with the wallclock time for
alanine dipeptide, whereas this is a linear scaling in the toy models. 
For example, an increase in $\tau$ from 100~fs to
1000~fs increases the wallclock time by a factor of two for alanine
dipeptide for WE simulations.

\section{Conclusions}

 We developed a steady--state path sampling procedure using the
weighted ensemble (WE) path sampling method. This procedure
does not depend on Markovian decomposition of the configuration
space and generated continuous paths from the starting to the
final state.
The steady state is established via a fairly simple extension of
the standard WE path sampling method --
a feedback loop into the starting state.
The ordinary rate (inverse mean--first--passage time) can
also be calculated.  A simple probability adjustment
procedure, the enhanced WE method, leads to 
a significantly more efficient attainment of steady state.
With an increase in the ruggedness of the energy landscape, the enhanced
WE method becomes more efficient as compared to both regular WE and
brute--force simulations.

 With minor changes to the probability
adjustment procedure, the enhanced WE method is
applicable for systems at equilibrium, and the probability
adjustment allows for a rapid equilibration.
We also suggest several possible improvements using optimal
bins and number of trajectories within a bin determined from initial paths,
as well as by hardwiring the weighted ensemble method into the
underlying dynamics code.

\clearpage

\section*{Appendix: Potential energies for toy systems}

\subsection*{One--dimensional system}

 The one--dimensional potential is constructed from six half wells
glued together from complementary pairs given by
\begin{equation}
U(x)=\frac{3h}{w^2}x^2\pm\frac{2h}{w^3}x^3
\label{ae1}
\end{equation}
where $h$ is the depth of each half well and $w$ is the (half) width.
The complementary pairs are given by the plus and the minus signs
that are valid in the $-w<x<0$ and $0<x<w$ ranges, respectively.
Such a potential leads to a smooth potential energy function.
In this work, $h=6k_{\mathrm{B}}T$ and $w=1$ m, and the minima are at
1 m, 3 m, and 5 m. The steep repulsions at $x<0$ and $x>6$ are modeled
by a $x^{12}$ potential, joined such that the potential energy function
remains smooth.

\subsection*{Two--dimensional system}

 The two--dimensional potential energy function is constructed by the
use of laying several energy wells on a flat surface. These energy 
wells are of the following form:
\begin{equation}
U(r)=-\left(\frac{2h}{w^2}r^2-\frac{h}{w^4}r^4\right)
\label{ae2}
\end{equation}
where $h$ and $w$ are the depth and the half width of the wells, respectively.
A negative value of $h$ indicates a minimum on the surface, whereas
a positive value indicates a maximum.
All wells width $w=0.5$~m.

 Table~\ref{t1} gives the values of $h$ used for generating the
smooth potential energy surface of Figure~\ref{2d_test1} 
for different values of $x$ and $y$.
Further, for $|x|>4.0$ or $|y|>4.0$, a repulsive potential of the
form: $100(x-4.0)^2$$kT$ is used (this is for $x>4.0$, and similar, forms
are used for other regions with absolute coordinate greater than 4.0).

 The rugged potential energy surface of Figure~\ref{2d_test} uses
the same form, and has
the same values of $w$ as above. The difference is in the values of $h$.
For this rugged potential, all positive values of $h$ are replaced by
10 $k_{\mathrm{B}}T$, and all
negative values of $h$ become $-5k_{\mathrm{B}}T$, 
except the one at (-3.5,-0.5)  for which
$h=-8k_{\mathrm{B}}T$.

\clearpage

\bfig
\resizebox{3.5in}{!}{\includegraphics{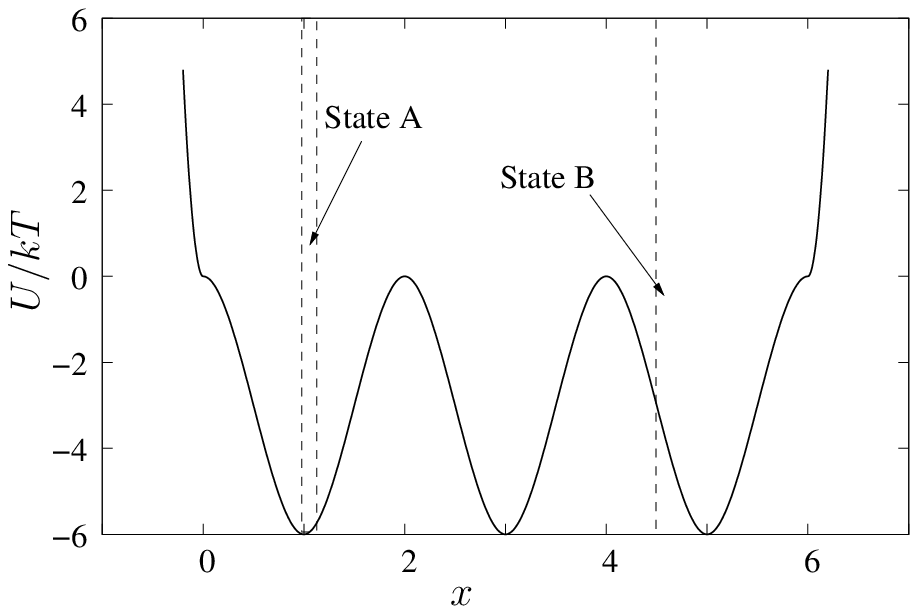}}\\
\caption{Potential energy profile for the one--dimensional
system. State B (target state) is defined by $x>4.5$, and
state A (initial state) is defined as $0.98<x<1.17$.}
\label{pot1d}
\efig

\clearpage

\bfig
\resizebox{3.5in}{!}{\includegraphics{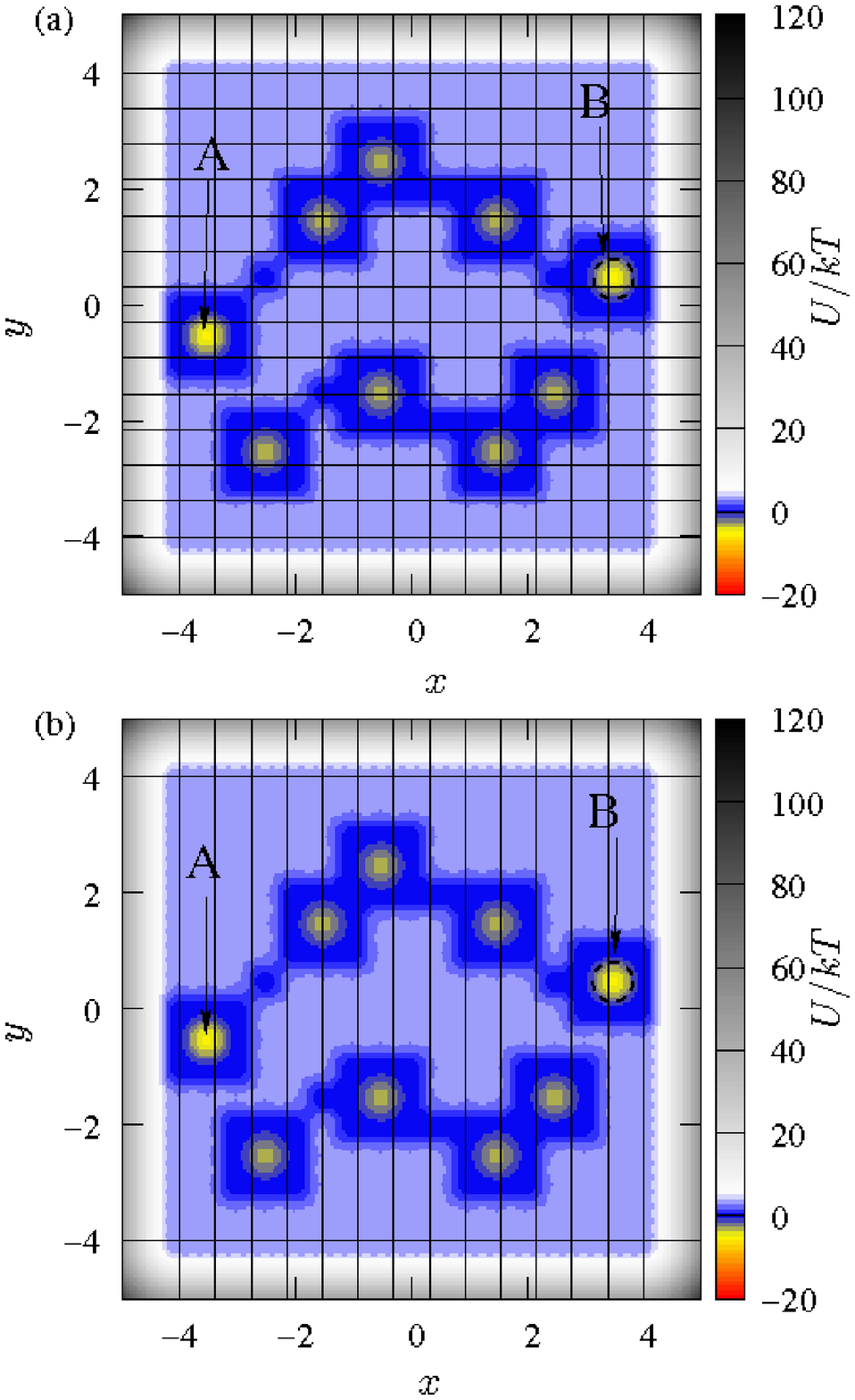}}\\
\caption{A relatively smooth two--dimensional landscape with two
``channels'' for transitions from A to B. Two different binning schemes
used in WE simulations are shown: (a) two dimensional (b) one dimensional.}
\label{2d_test1}
\efig

\clearpage

\bfig
\resizebox{3.5in}{!}{\includegraphics{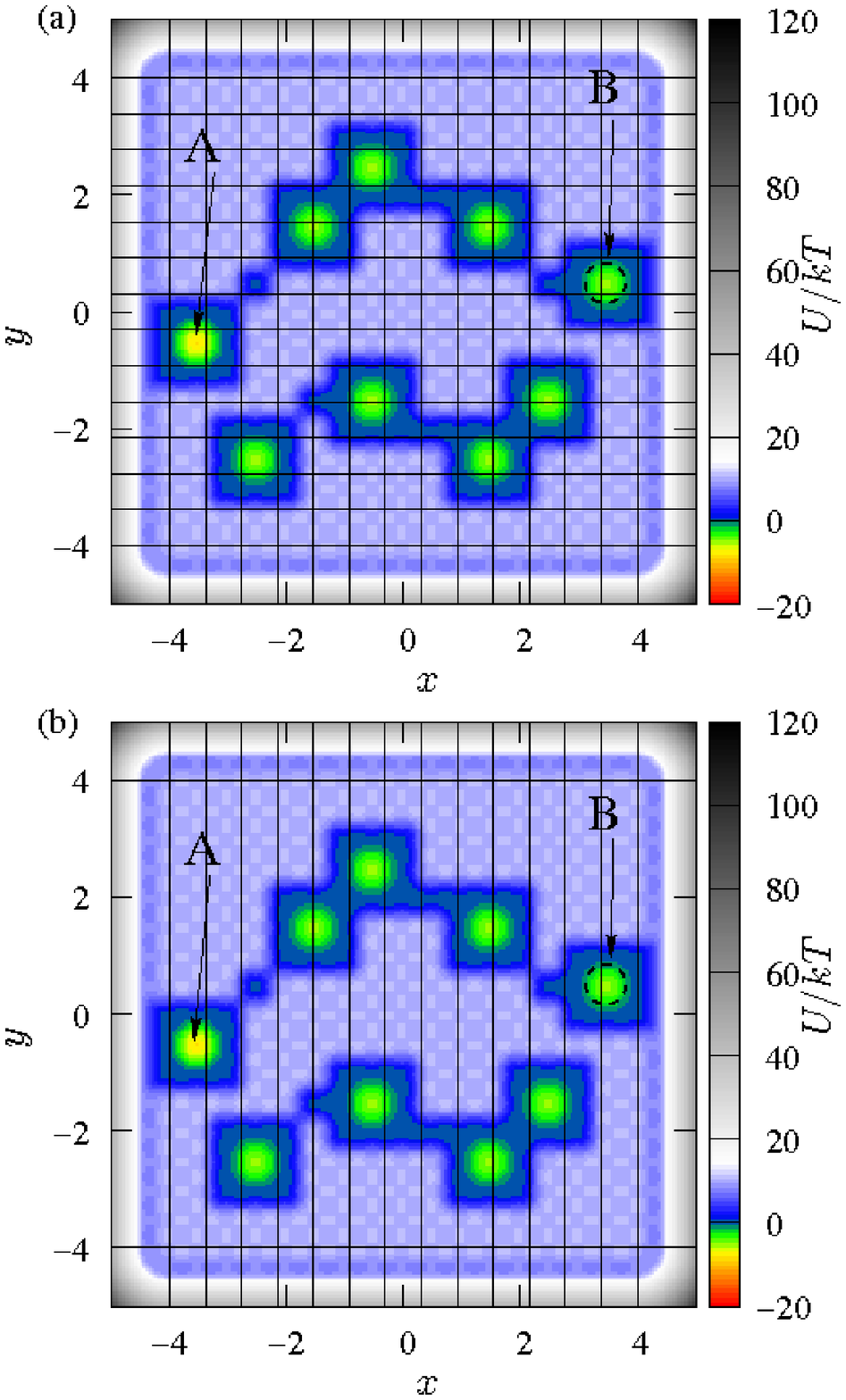}}\\
\caption{A relatively rugged two--dimensional landscape with two
``channels'' for transitions from A to B. Two different binning schemes
used in WE simulations are shown: (a) two dimensional (b) one dimensional.}
\label{2d_test}
\efig

\clearpage

\bfig
\resizebox{3.5in}{!}{\includegraphics{AlaD-Psi-Phi.epsi}}\\
\caption{The $\Psi$--$\Phi$ plane for alanine dipeptide. Red dots were
obtained from a long brute--force simulation. 
The initial state is contained in C$_{7\mathrm{eq}}$ and the
target state is C$_{7\mathrm{ax}}$ as indicated by the circles. 
Also shown via circles are the right--
and left--handed helical regions, $\mathrm{\alpha_R}$ and
$\mathrm{\alpha_L}$, respectively. Grid lines represent the two--dimensional
binning.}
\label{adp_pot}
\efig

\clearpage

\bfig
\resizebox{3.5in}{!}{\includegraphics{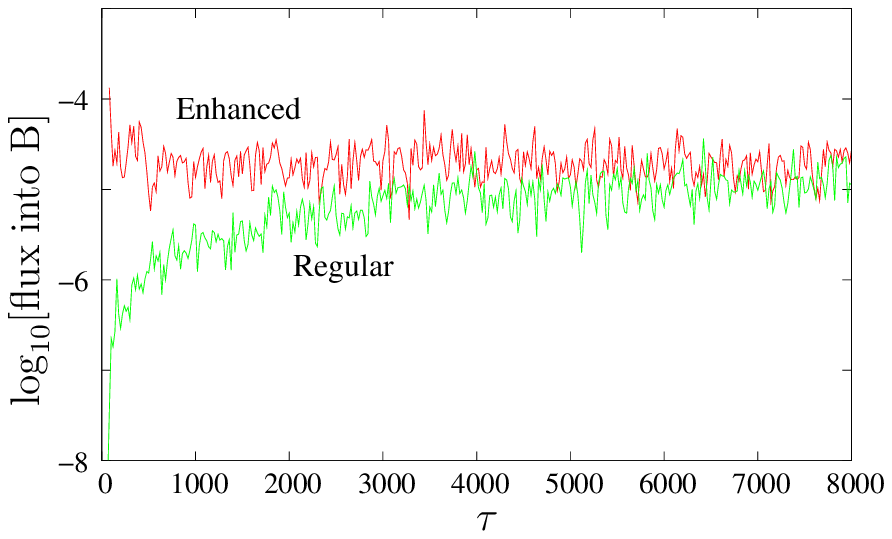}}\\
\caption{A comparison of fluxes (on log scale) into state B obtained 
via enhanced (black 
line) and regular (red line)  WE implementation of steady--state procedure for
the one--dimensional system of Figure~\ref{pot1d}.
Flux into state B for the one--dimensional system. Due to the probability
adjustment procedure, the enhanced version reaches steady state significantly
faster. All results shown are window averaged over 100 time increments, $\tau$.}
\label{flux1d}
\efig

\clearpage

\bfig
\begin{tabular}{c}
\resizebox{3.5in}{!}{\includegraphics{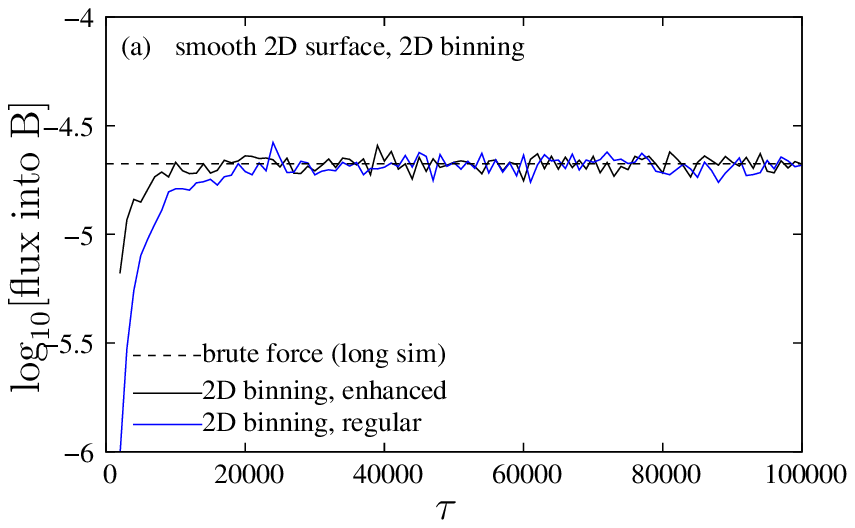}}\\
\resizebox{3.5in}{!}{\includegraphics{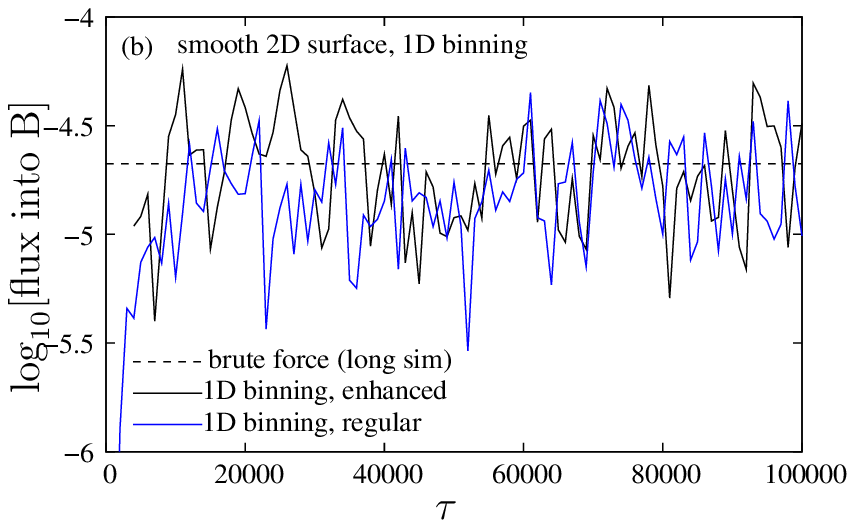}}\\
\end{tabular}
\caption{Comparison of WE methods and binning approaches for a 
smooth two--dimensional potential energy surface.
Results from 2D binning are shown in panel (a) and from 1D WE binning 
are in panel (b). The flux reaches a steady value sooner for the enhanced 
version than for the regular version. All results agree with each other and 
also with the final result from long brute--force simulations. All results 
shown are window averaged using 100 $\tau$ windows.}
\label{flux2d_4}
\efig

\clearpage

\bfig
\begin{tabular}{c}
\resizebox{3.5in}{!}{\includegraphics{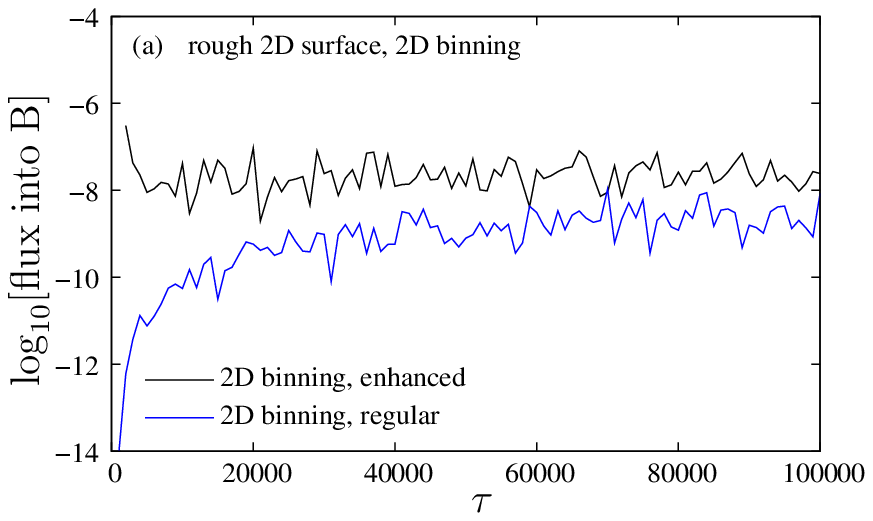}}\\
\resizebox{3.5in}{!}{\includegraphics{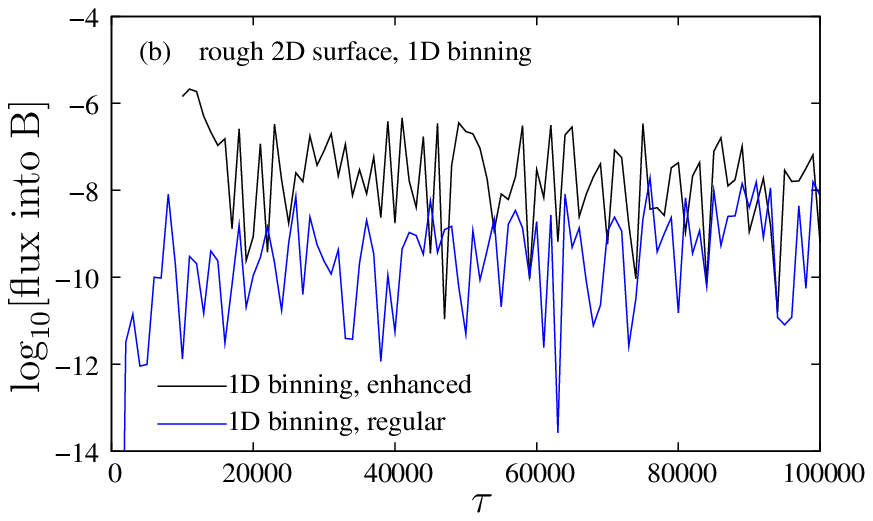}}\\
\end{tabular}
\caption{Comparison of WE methods and binning approaches for a 
rugged two--dimensional potential energy surface.
Results from 2D binning are shown in panel (a) and from 1D WE binning 
are in panel (b). The flux reaches a steady value sooner for the enhanced 
version than for the regular version. All results agree with each other and 
also with the final result from long brute--force simulations. All results 
shown are window averaged using 100 $\tau$ windows.}
\label{flux2d}
\efig

\clearpage

\bfig
\resizebox{3.5in}{!}{\includegraphics{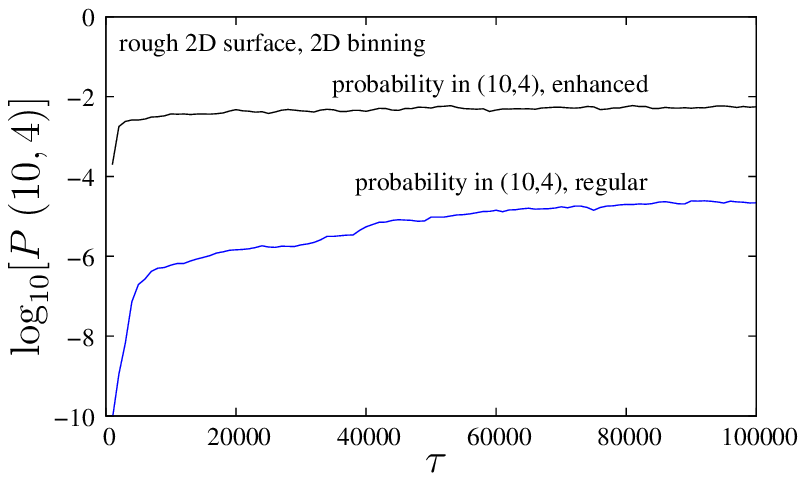}}\\
\caption{Comparison of WE methods for a single intermediate.
Probabilities (log scale) in the bin with indices 
given by (10,4) are plotted {\it versus} time for regular and
enhanced WE simulations.
The enhanced version reaches a steady value
fairly quickly, whereas the regular version does not. All results are
window averaged using 100 $\tau$ windows.}
\label{pb2d}
\efig

\clearpage

\bfig
\resizebox{3.5in}{!}{\includegraphics{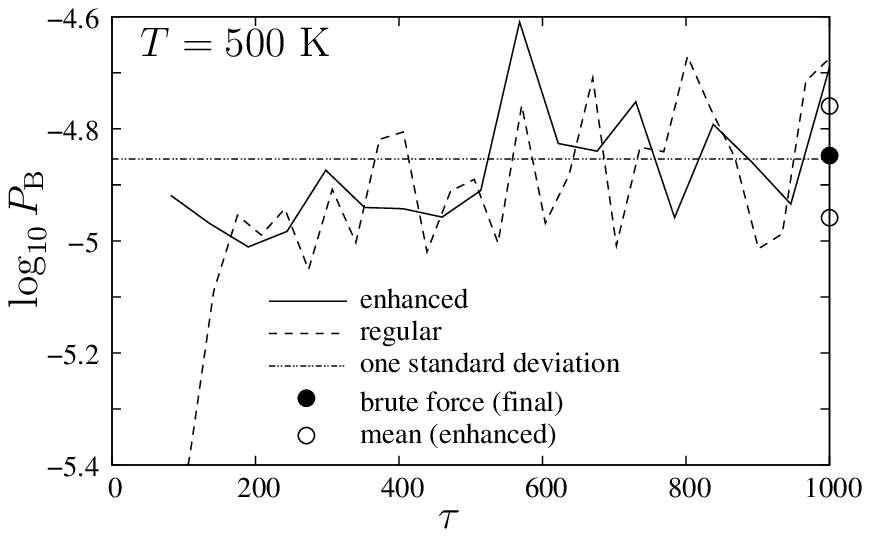}}\\
\caption{Rate estimation for alanine dipeptide at 500~K. The
flux into the C$_{7\mathrm{ax}}$ is plotted based on the
enhanced WE, regular WE, and final results from long
brute--force simulations. All results converge to the same
final value, and the enhanced WE is not more efficient than
regular WE simulation at this elevated temperature. 
All results are window averaged using 50 $\tau$ windows.}
\label{adp_500}
\efig

\clearpage

\bfig
\begin{tabular}{c}
\resizebox{3.5in}{!}{\includegraphics{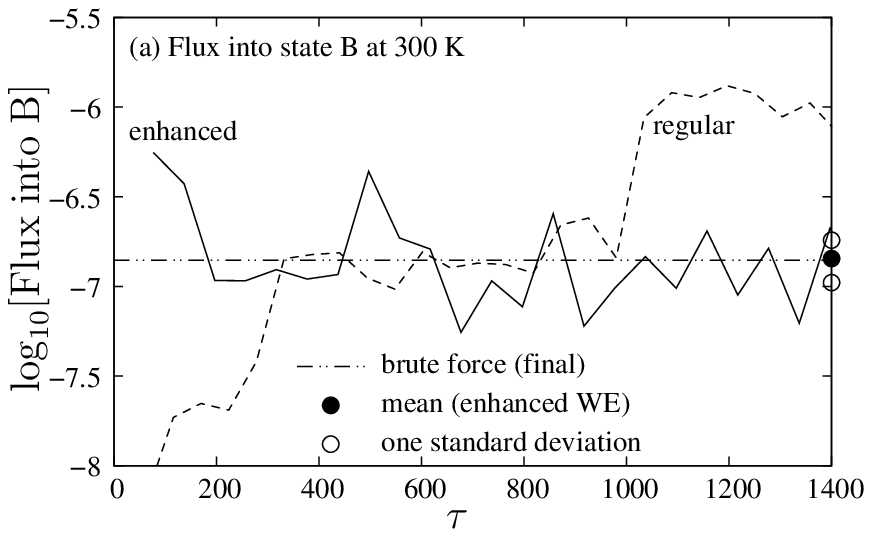}}\\
\resizebox{3.5in}{!}{\includegraphics{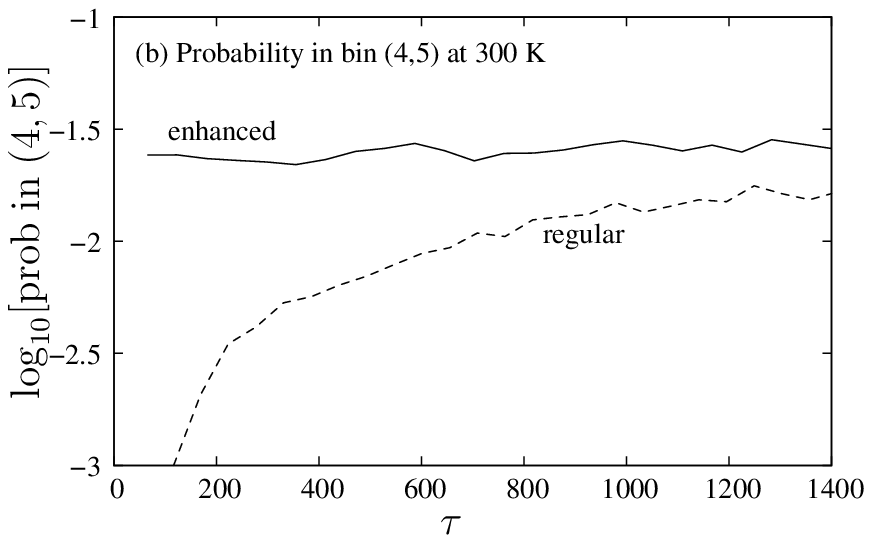}}\\
\end{tabular}
\caption{Comparison of enhanced and regular WE methods for
alanine dipeptide at 300~K.
The steady state flux into the C$_{7\mathrm{ax}}$ state 
via the two WE methods and brute--force simulations is shown
in panel (a), and the probability in the bin (4,5) in the
$\alpha_{\mathrm{R}}$ intermediate state is shown in panel (b).
Both panels (a) and (b) show that the regular WE simulation is
unable to reach steady state, whereas steady state
is established rapidly with the enhanced WE method.
All results are window averaged using 50 $\tau$ windows.}
\label{adp_300}
\efig

\clearpage

\bfig
\resizebox{3.5in}{!}{\includegraphics{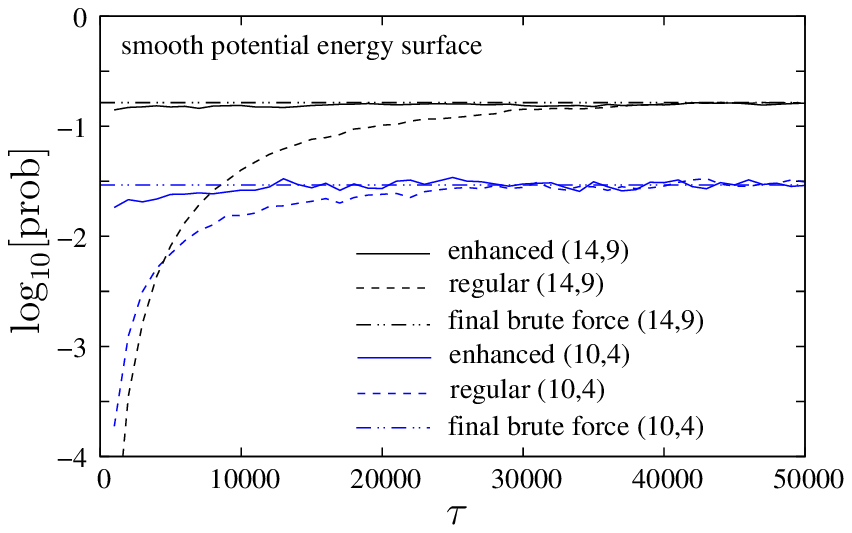}}\\
\caption{Equilibrium sampling via WE simulation for the smooth
two--dimensional potential energy surface. The plot compares
probabilities obtained in two different bins (indices (14,9) and (10,4)
of Figure~\ref{2d_test1} (a)) for
equilibrium simulations attempted using regular and enhanced WE simulations.
Final results from a long brute--force simulation are also shown for
reference. The bin probabilities in the enhanced version reach equilibrium
values significantly quicker than the regular version.}
\label{equil}
\efig

\clearpage

\bfig
\resizebox{3.5in}{!}{\includegraphics{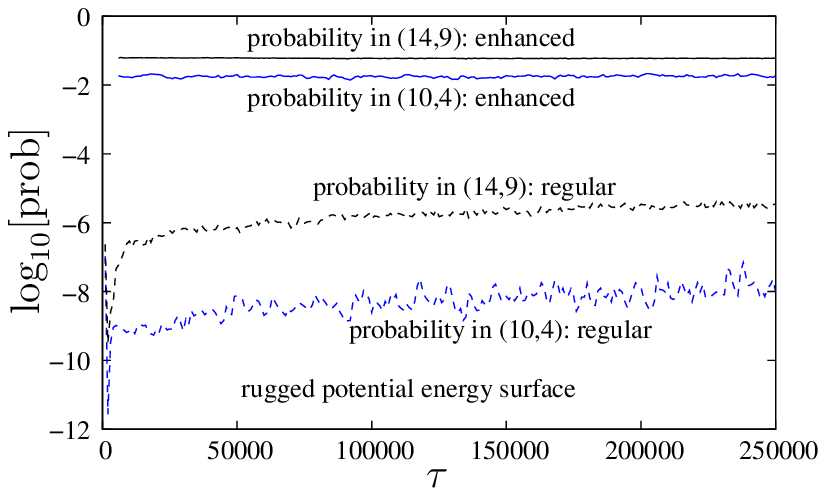}}\\
\caption{Equilibrium sampling via WE simulation for the rugged
two--dimensional potential energy surface. The plot compares
probabilities obtained in two different bins (indices (14,9) and (10,4)
of Figure~\ref{2d_test} (a)) for
equilibrium simulations attempted using regular and enhanced WE simulations.
The bin probabilities in the enhanced WE simulations reach equilibrium
fairly rapidly, whereas the regular WE simulations do not reach
equilibrium in the timescale of the simulation.}
\label{equil_pot2}
\efig

\clearpage

\begin{table}
\begin{center}
\caption{Well depths, $h$, in units of $k_{\mathrm{B}}T$ for the smooth
potential energy surface. Negative values indicate minima on the
surface, whereas positive values indicate  maxima.}
\begin{tabular}{c|c|c|c|c|c|c|c|c|c|c|c|c|c|c|c|c|c}
     & -4 & -3.5 & -3 & -2.5 & -2 & -1.5 & -1 & -0.5 & 0 & 0.5 & 1 & 1.5 &  2 & 2.5 & 3 & 3.5 & 4 \\ \hline
-4   & 3  & 3    & 3  & 3    & 3  & 3    & 3  &  3   & 3 & 3   & 3 & 3   & 3  & 3   & 3 & 3   & 3\\ \hline
-3.5 & 3  & 3    & 3  & 3    & 3  & 3    & 3  &  3   & 3 & 3   & 3 & 3   & 3  & 3   & 3 & 3   & 3\\ \hline
-3   & 3  & 3    & 0  & 0    & 0  & 3    & 3  &  3   & 3 & 3   & 0 & 0   & 0  & 3   & 3 & 3   & 3\\ \hline
-2.5 & 3  & 3    & 0  &-3    & 0  & 3    & 3  &  3   & 3 & 3   & 0 &-3   & 0  & 3   & 3 & 3   & 3\\ \hline
-2   & 3  & 3    & 0  & 0    & 0  & 3    & 0  &  0   & 0 & 0   & 0 & 0   & 0  & 0   & 0 & 3   & 3\\ \hline
-1.5 & 3  & 3    & 3  & 3    & 3  & 0    & 0  & -3   & 0 & 3   & 3 & 3   & 0  &-3   & 0 & 3   & 3\\ \hline
-1   & 0  & 0    & 0  & 3    & 3  & 3    & 0  &  0   & 0 & 3   & 3 & 3   & 0  & 0   & 0 & 3   & 3\\ \hline
-0.5 & 0  &-5    & 0  & 3    & 3  & 3    & 3  &  3   & 3 & 3   & 3 & 3   & 3  & 3   & 3 & 3   & 3\\ \hline
 0   & 0  & 0    & 0  & 3    & 3  & 3    & 3  &  3   & 3 & 3   & 3 & 3   & 3  & 3   & 0 & 0   & 0\\ \hline
 0.5 & 3  & 3    & 3  & 0    & 3  & 3    & 3  &  3   & 3 & 3   & 3 & 3   & 3  & 0   & 0 &-5   & 0\\ \hline
 1   & 3  & 3    & 3  & 3    & 0  & 0    & 0  &  3   & 3 & 3   & 0 & 0   & 0  & 3   & 0 & 0   & 0\\ \hline
 1.5 & 3  & 3    & 3  & 3    & 0  &-3    & 0  &  3   & 3 & 3   & 0 &-3   & 0  & 3   & 3 & 3   & 3\\ \hline
 2   & 3  & 3    & 3  & 3    & 0  & 0    & 0  &  0   & 0 & 0   & 0 & 0   & 0  & 3   & 3 & 3   & 3\\ \hline
 2.5 & 3  & 3    & 3  & 3    & 3  & 3    & 0  & -3   & 0 & 3   & 3 & 3   & 3  & 3   & 3 & 3   & 3\\ \hline
 3   & 3  & 3    & 3  & 3    & 3  & 3    & 0  &  0   & 0 & 3   & 3 & 3   & 3  & 3   & 3 & 3   & 3\\ \hline
 3.5 & 3  & 3    & 3  & 3    & 3  & 3    & 3  &  3   & 3 & 3   & 3 & 3   & 3  & 3   & 3 & 3   & 3\\ \hline
 4   & 3  & 3    & 3  & 3    & 3  & 3    & 3  &  3   & 3 & 3   & 3 & 3   & 3  & 3   & 3 & 3   & 3\\ 
\end{tabular}
\label{t1}
\end{center}
\end{table}

\end{document}